\documentclass[aps,prl,showpacs,showkeys,twocolumn,usletter,groupedaddress]{revtex4}

\usepackage{amsmath}
\usepackage{amsfonts}
\usepackage{amssymb}
\usepackage{latexsym}
\usepackage{graphicx}
\usepackage{float}
\usepackage{wrapfig}
\usepackage{epsfig}

\newcommand{\be}{\begin{eqnarray}}
\newcommand{\ee}{\end{eqnarray}}

\newcommand{\bmat}{\left ( \begin{array}{cc} 
}
\newcommand{\emat}{\end{array} \right )}

\newcommand{\U}{\textrm{U}}

\newcommand{\tr}{\textrm{tr}}

\newcommand{\diag}{\textrm{diag}}

\newcommand{\eins}{\leavevmode\hbox{\small1\kern-3.8pt\normalsize1}}

\begin{document}
 \title{On the Eigenvalue Density of the non-Hermitian Wilson Dirac Operator}
\author{Mario Kieburg, Jacobus J.M. Verbaarschot, and Savvas Zafeiropoulos}
\date{\today}
\affiliation{Department of Physics and Astronomy, State University of New York at Stony Brook, NY 11794-3800, USA}

\begin{abstract}
 We find the lattice spacing dependence
of the eigenvalue density of the non-Hermitian Wilson Dirac operator in the 
$\epsilon$-domain.
The starting point is the joint probability density of the corresponding
random matrix theory. 
In addition to the density of the complex eigenvalues we also obtain the
density of the real eigenvalues separately for positive and negative
chiralities as well as an explicit analytical expression for the 
number of additional real modes.
\end{abstract}

\maketitle

\paragraph{Introduction.}
In the past two decades, there has been an increasing interest in non-Hermitian random matrix theory (RMT) \cite{RMT1}. To name a few applications, quantum chaos in open systems \cite{SSS}, dissipative systems \cite{DIS} and QCD at finite chemical potential \cite{ST}. Some features of the model we are considering also occur in the condensed matter system analyzed in Ref.~\cite{RL}.

The connection between the infrared limit of QCD and RMT
 has been  well understood in the continuum limit since the early 90's~\cite{ShuVer93}.
It is based on the universality
of chiral RMT in the microscopic limit (or $\epsilon$ domain) \cite{ADMN97}
with  chiral RMT described by the same chiral Lagrangian as QCD.
The main advantage of RMT is the availability of powerful methods to derive
analytical results, and  recently  this approach was applied to
QCD at finite lattice spacing \cite{osborn,DSV10,ADSV10b,SplVer11}.
 It was shown that the $\epsilon$ limit of the
 chiral Lagrangian for the Wilson Dirac 
operator $D_{\rm W}$ \cite{ShaSin98,BRS04} 
can be obtained from an equivalent RMT.
Discretization effects 
of the spectrum of $D_{\rm W}$ have been studied directly by means of chiral 
Lagrangians~\cite{Sha06,necco,DSV10,ADSV10b,SplVer11}, but using RMT
methods will enable us to obtain results that were not accessible
previously.

The aim of this paper is to obtain  analytical results for 
the eigenvalue density of $D_{\rm W}$ for the RMT model proposed 
in Ref.~\cite{DSV10}. We consider the quenched case.


\paragraph{RMT.}

We consider the random matrix theory~\cite{DSV10},
\begin{eqnarray}\label{1}
 D_{\rm W}&=&\left(\begin{array}{cc} aA & W \\ -W^\dagger & aB\end{array}\right)
\end{eqnarray}
distributed by
\begin{eqnarray}\label{2}
 P(D_{\rm W})\propto\exp\left[-\frac{n}{2}(\tr A^2+\tr B^2)-n\tr WW^\dagger\right].
\end{eqnarray}
The matrices $A$ and $B$ are Hermitian $n\times n$ and $(n+\nu)\times(n+\nu)$ 
matrices, respectively, and the entries of $W$ are complex and independent. 
In the microscopic limit, with  $n\to\infty$ at fixed rescaled  eigenvalues
$\widehat{z}=2nz$ and  lattice spacing $\widehat{a}^2=na^2/2$, the spectral properties of
this RMT become universal and agree with Wilson chiral perturbation theory in the same limit (with $n$ identified
as the volume of space-time) apart from the squared trace terms~\cite{BRS04,Sha06}.
 The finite integer $|\nu|\leq n$ is the index of the Dirac operator and  is
kept fixed.

The matrix $D_{\rm W}$ is $\gamma_5=\diag(\eins_n,-\eins_{n+\nu})$-Hermitian, i.e. $D_{\rm W}^\dagger=\gamma_5D_{\rm W}\gamma_5$. Therefore its eigenvalues are
either real or come in complex conjugate pairs.
 The $\nu$ generic zero modes at $a=0$ become the generic 
real modes of $D_{\rm W}$ at finite lattice spacing.
Furthermore, $D_{\rm W}$ may have $2n-2l $ additional real eigenvalues which appear
when a pair of complex eigenvalues collides with the real axis.

In Refs.~\cite{DSV10,ADSV10b,SplVer11,AkeNag11} the 
technically simpler case of the Hermitian 
Wilson Dirac operator $D_5=\gamma_5D_{\rm W}$ was studied. Although
 spectra of $D_5$ have been studied in the lattice literature
\cite{Luscher}, only 
 the eigenvalues of $D_{\rm W}$  are directly
related to chiral symmetry breaking which is our main motivation to
study its spectral properties.

\paragraph{The joint probability distribution (jpd).}
To preserve the $\gamma_5$-Hermiticity of $D_{\rm W}$ we can only quasi-diagonalize $D_{\rm W}$ by
a non-compact unitary matrix $ U \in \U(n,n+\nu)$,
\begin{eqnarray}\label{3}
 D_{\rm W}=UXU^{-1}\ \textrm{and}\ X=\left(\begin{array}{cccc} x_1 & 0 & 0 & 0 \\ 0 & x_2 & y_2 & 0 \\ 0 & -y_2 & x_2 & 0 \\ 0 & 0 & 0 & x_3 \end{array}\right).
\end{eqnarray}
In contrast to the  diagonalization of a Hermitian matrix
such as $D_5$, the matrix $X$ may only be quasi-diagonal where 
$x_1$, $x_2$, $y_2$ and $x_3$ are diagonal matrices of dimension 
$n-l$, $l$, $l$ and $n-l+\nu$ with $0\leq l\leq n$ the number of complex 
conjugate pairs. The complex eigenvalues are given by 
$(z,z^*)=(x_2+\imath y_2,x_2- \imath y_2)$. The ensemble $D_{\rm W}$ decomposes into 
$n+1$ disjoint
sets of quasi-diagonal matrices~\eqref{3} with a fixed number of real 
eigenvalues. 
The joint probability density  of the $2n+\nu$ 
eigenvalues $Z=(z_{1{\rm r}},\ldots,z_{n{\rm r}},z_{1{\rm l}},\ldots,z_{n+\nu,{\rm l}})\in\mathbb{C}^{2n+\nu}$ 
can be obtained by integrating over $U$. 
This calculation will be discussed
in detail elsewhere. We only give the result for $\nu\geq0$ which is not a 
restriction because of the symmetry $\nu\rightarrow-\nu$. The jpd is given by
\begin{eqnarray}
 &&p(Z)\propto\Delta_{2n+\nu}(Z)\det\left[\begin{array}{c}  \left\{g_2(z_{a{\rm r}},z_{b{\rm l}})\right\}\underset{1\leq b\leq n+\nu}{\underset{1\leq a\leq n}{\ }} \\ \left\{z_{b{\rm l}}^{a-1}g_1(z_{b{\rm l}})\right\}\underset{1\leq b\leq n+\nu}{\underset{1\leq a\leq \nu}{\ }} \end{array}\right],\nonumber\\
 &&g_1(z)=\sqrt{\frac{n}{2\pi a^2}}\exp\left[-\frac{n}{2a^2}x^2\right]\delta(y),\label{4}
\end{eqnarray}
\begin{eqnarray*}
 &&g_2(z_1,z_2)=
\sqrt{\frac{n^3}{4\pi a^2(1+a^2)}}\frac{z_1^*-z_2^*}{|z_1-z_2|}\nonumber\\
 &\times&\left[\exp\left[-\frac{n(x_1+x_2)^2}{4a^2}-\frac{n(y_1-y_2)^2}{4}\right]
\delta^{(2)}(z_1-z_2^*)
\right.\nonumber\\
 &+&\left.\frac{1}{2}\exp\left[-\frac{n}{4a^2}(x_1+x_2)^2+\frac{n}{4}(x_1-x_2)^2\right]\right.\nonumber\\
 &\times&\left.{\rm erfc}\left[\frac{\sqrt{n(1+a^2)}}{2a}|x_1-x_2|\right]\delta(y_1)\delta(y_2)\right],\nonumber\\
&&\equiv g_{2\,{\rm  c}}(z_1)\delta^{(2)}(z_1-z_2^*)+
g_{2\,{\rm  r}}(x_1,x_2)\delta(y_1)\delta(y_2),
\end{eqnarray*}
\vspace*{-1.1cm}\begin{center}
\begin{figure}[h!]
\includegraphics[scale=0.345]{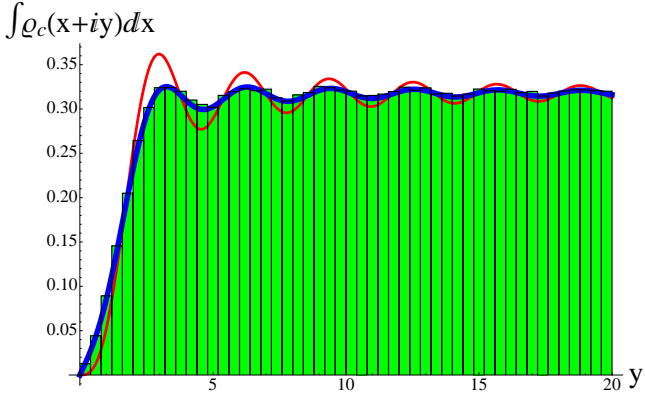}
\caption{The projection of $\rho_{\rm c}$ onto the imaginary axis for $\nu=1$ and $\widehat{a}=1/\sqrt{8}$. The Monte Carlo simulation (histogram, bin size=0.4) contains 200000 matrices with $n=50$. This simulation nicely confirms our analytical result (blue curve) and shows the deviations from the $\widehat{a}=0$ result (red curve).}
\label{fig1}
\end{figure}
\end{center}
\vspace*{-1.1cm}where ${\rm erfc}$ is the complementary error function and
$\delta^{(2)}(x+\imath y)=\delta(x)\delta(y)$. 
Due to $\gamma_5$ the permutation group $\mathfrak{S}(2n+\nu)$ 
is broken to $\mathfrak{S}(n)\times\mathfrak{S}(n+\nu)$ 
which reflects itself in the product of the 
Vandermonde determinant $\Delta_{2n+\nu}(Z)$ and the other 
determinant in Eq.~\eqref{4}. The expansion of the delta functions  
yields the jpd  for each of the $n+1$ subsets with a fixed number of 
complex eigenvalue pairs. The two-point distribution $g_2$ splits into one term for the real eigenvalues $g_{2{\rm r}}$ and one for the complex conjugated pairs $g_{2{\rm c}}$ as it is also known for the real Ginibre ensemble and its chiral counterpart \cite{Gin}.

\paragraph{The eigenvalue densities} for the real and complex eigenvalues
 can be obtained by integrating over all eigenvalues except one. The spectral
density 
can be decomposed into the density of real modes, $\rho_{\rm r}$ 
for positive chirality ($\langle\psi|\gamma_5|\psi\rangle>0$), $\rho_{\rm l}$ for negative chirality ($\langle\psi|\gamma_5|\psi\rangle<0$), 
and the density 
of complex pairs, $\rho_{\rm c}$,
\vspace*{-0.5cm}
\begin{eqnarray}
 \int p(Z)\prod_{z_j\neq z_{1{\rm r}}}d[z_j]
&=&\rho_{\rm r}(x_{1{\rm r}})\delta(y_{1{\rm r}})+\frac{\rho_{\rm c}(z_{1{\rm r}})}{2} \label{5},\\
 \int p(Z)\prod_{z_j\neq z_{1{\rm l}}}d[z_j]
&=&\rho_{\rm l}(x_{1{\rm l}})\delta(y_{1{\rm l}})+\frac{\rho_{\rm c}(z_{1{\rm l}})}{2}\label{6}.
\end{eqnarray}
Note that the chirality reflects the conventions of the RMT.
By expanding the first row of the determinant in Eq.~(\ref{4}) 
and re-expressing the additional factors from $\Delta_{2n+\nu}(Z)$ as $N_{\rm f} =2$ partition functions we obtain
\begin{eqnarray}
\rho_{\rm c}(z)& = &g_{2{\rm c}}(z)(z-z^*) Z_{N_{\rm f}=2}^\nu(z,z^*;a), \label{7a}\\
\rho_{\rm r}(x) &=& \int_{\mathbb{R}} g_{2{\rm r}}(x,x^\prime)(x-x^\prime) Z_{N_{\rm f}=2}^\nu(x,x^\prime;a) dx^\prime. \label{7b}
\end{eqnarray}
A similar factorized structure was found in Ref.~\cite{Splittorff:2003cu}.

Expanding the first column of the determinant and integrating  over
all eigenvalues except $z_{1l} $, we find the same expression for $\rho_{\rm c}$ and the density $\rho_{\rm l}$ of the real modes
originating from $g_2$ (using  Eq.~(6)). However, there is an additional contribution to $\rho_{\rm r}$ due
to the last $\nu $ rows which gives the distribution of chirality over the real eigenvalues
\be
\rho_\chi = \rho_{\rm l} -\rho_{\rm r}. 
\ee
Additional rows of some of the determinants have to be expanded
 to express them into known partition functions. 
We have checked for $\nu =1$ and $\nu =2 $ that 
the result agrees with previously derived
expressions \cite{DSV10,SplVer11}.

In the microscopic limit, the partition functions in Eqs.~(\ref{7a}) and (\ref{7b}) can be expressed in terms of 
integrals over $\U(2)$. They can be simplified using the
eigenvalues of the $\U(2)$-matrices as integration variables resulting in
\begin{widetext}
\begin{eqnarray}
 \rho_{\rm c}\left(\frac{z}{2n}\right)&=&\frac{e^{-x^2/8\widehat a^2}|y|}{(2\pi)^{5/2}2\widehat{a}}
\int_{[0,2\pi]^2} e^{x(\cos \varphi_1 +\cos \varphi_2)- 4 \widehat a^2(\cos^2 \varphi_1+\cos^2 \varphi_2)} 
{\rm sinc}[y(\cos\varphi_1-\cos\varphi_2)]\cos\nu(\varphi_1+\varphi_2) D\varphi_k,
\label{8}\\
 \rho_{\rm r}\left(\frac{x}{2n}\right)&=&\frac{1}{16\pi^2}
\int_{[0,2\pi]^2}\frac{ \exp[\Delta_1^2-\Delta_2^2]{\rm erf}\left[\Delta_1,\sqrt{2}\Delta_1\right]-\exp[\Delta_2^2-\Delta_1^2]{\rm erf}\left[\Delta_2,\sqrt{2}\Delta_2\right]}{\cos\varphi_1-\cos\varphi_2} \cos\nu(\varphi_1+\varphi_2) D\varphi_k,
\\
 \rho_\chi\left(\frac{x}{2n}\right)&=&\frac{(-1)^\nu}{16\pi \widehat{a}^2}\int_{\mathbb{R}^2} \frac{e^{-((s_1-x)^2+(s_2+ix)^2)/16\widehat{a}^2}}{s_1-\imath s_2}s_1^{\nu}
\left[s_1  K_{\nu+1}(s_1)I_{\nu}(\imath s_2) +\imath s_2  K_{\nu}(s_1)I_{\nu+1}(\imath s_2)\right]\frac{\delta^{(\nu-1)}(s_1)}{(\nu-1)!}ds_1ds_2.\label{10}
\end{eqnarray}
\end{widetext}
The functions ${\rm sinc}$, ${\rm erf}$, $I_l$, $ K_l$ and $\delta^{(l)}$ are
 the \textit{sinus cardinalis}, the generalized incomplete error function\newpage
\begin{center}
\begin{figure}[t!]
\includegraphics[scale=0.3]{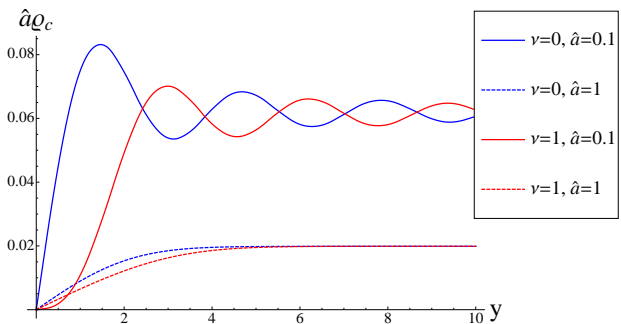}
\caption{Along the imaginary axis the difference of $\rho_{\rm c}$ for different $\nu$ is much clearer at small $\widehat{a}$ (solid curves) than at large $\widehat{a}$ (dashed curves) where they are almost the same.}
\label{fig5}
\end{figure}
\end{center}
\vspace*{-1.0cm}
(${\rm erf}(b,c)={\rm erf}(c)-{\rm erf}(b)$), 
modified Bessel function of the first and second kind and the $l$-th derivative of the
Dirac delta function, respectively. 
The integration measure is induced by the
invariant $ \U(2)$ measure, $D\varphi_k =$
$\sin^2((\varphi_1 -\varphi_2)/2)d\varphi_1d \varphi_2$ and $\Delta_j=2\widehat{a}\left(\cos\varphi_j-{x}/{8\widehat{a}^2}\right)$.
Because of the $\delta$-function only the algebraic singular part of the 
$K_\nu$ contributes to $\rho_\chi$ (which was already obtained in Refs.~\cite{DSV10,SplVer11}).
The distribution $\rho_\chi$ vanishes for $\nu=0$ and can be obtained from the generating function for the eigenvalue
density of $\gamma_5(D_W +m)$.~\cite{DSV10} Comparisons of the analytical results with simulations of the random matrix model~\eqref{1} are shown in Figs.~\ref{fig1} and \ref{fig2}. 
The normalizations are chosen such that the integral over $\rho_\chi$ is equal to $\nu$. 
The other constants are already fixed by this choice.

\begin{center}
\begin{figure}[t!]
\includegraphics[scale=0.35]{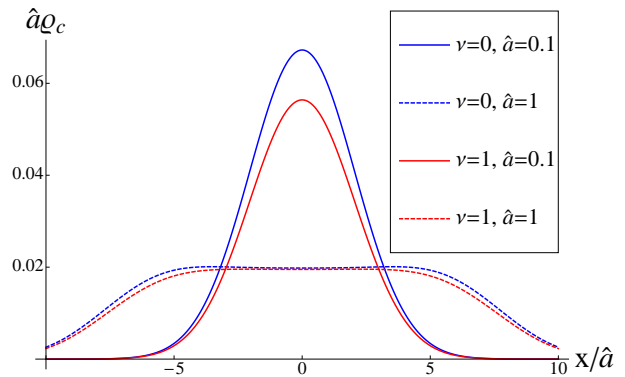}
\caption{The distribution $\rho_{\rm c}$ along a parallel axis to the x-axis (here at $y=40\widehat{a}^2$) is Gaussian shaped for small $\widehat{a}$ (solid curves) and develops a plateau for large lattice spacing (dashed curves).}
\label{fig6}
\end{figure}
\end{center}

\begin{center}
  \begin{figure}[b!]
\includegraphics[scale=0.54]{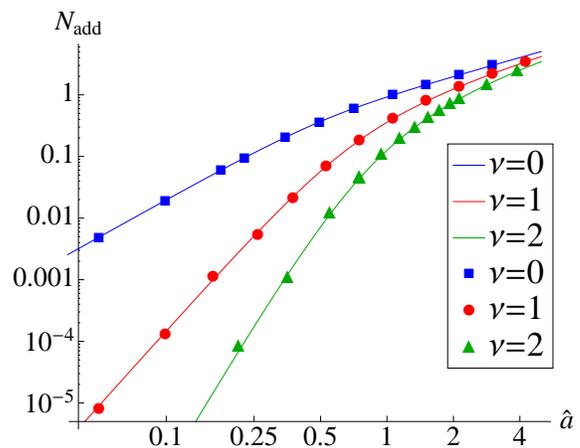}
\caption{Log-log-plot of the additional real eigenvalues versus $\widehat{a}$ 
for various $\nu$. The number of matrices and its size vary 
in this plot for the Monte Carlo simulations (symbols). 
The statistical error of the numerics varies between $0.1\%$ and $10\%$ around the analytic result (solid curves).}
\label{fig2}
\end{figure} 
\end{center}

For small increasing $\widehat{a}$ the complex eigenvalues move parallel to the
real axis according to a Gaussian distribution with a width of $2\widehat{a}$ (See Fig.~\ref{fig6}). 
Therefore the density of the projection of these eigenvalues on the
imaginary axis is very close to the $\widehat{a} = 0$ result. 
For large  $\widehat{a}$ the distribution of the real parts
of the complex eigenvalues 
 develops a box-like shape   
from $-8\widehat{a}^2$ to $8\widehat{a}^2$ which can be derived from
a saddle point 
approximation of Eq.~(\ref{8})  (See Fig.~\ref{fig6}) and the oscillations
disappear (See Fig.~\ref{fig5}). Along the imaginary  axis $\rho_{\rm c}(\widehat{a}\gg1)$ becomes $\widehat{a}^{-2}{\rm erf}(y/4\widehat{a})$.

Near the real axis  $\rho_{\rm c}$ behaves as $y^{\nu+1}$ for small $\widehat{a}$ but is linear in $y$ for $\widehat{a}$ large enough (See Fig.~\ref{fig5}).
 In the continuum limit it
 peaks around the imaginary axis and
 eventually gets the form of the continuum microscopic eigenvalue density. 

\paragraph{Real modes.}
For a Wilson Dirac operator with index $\nu$ there are at least $\nu$ real modes.
The additional real modes result when complex conjugate
 eigenvalue pairs enter the real axis. The average number of
these modes follows from the integral
 \be 
 N_{\rm add}=2\int_{\mathbb{R}} \rho_{\rm r}\left(\frac{x}{2n}\right)dx\hspace*{3cm} \label{11}\\
 =\int_{[0,2\pi]^2}\frac{1-e^{-2\widehat{a}^2(\cos\varphi_1-\cos\varphi_2)^2}}{8\pi^2\sin^2((\varphi_1+\varphi_2)/2)}
\cos\nu(\varphi_1+\varphi_2)d\varphi_1d\varphi_2.\nonumber
\ee
 In the limits for small and large lattice spacing we find
\begin{eqnarray}\label{12}
 N_{\rm add}\propto\left\{\begin{array}{cl} \widehat{a}^{2(\nu+1)}, & \widehat{a}\ll1,\\ \widehat{a}, & \widehat{a}\gg1.\end{array}\right.
\end{eqnarray}
\vspace*{-1.4cm}This is shown in Fig.~\ref{fig2}. For large lattice spacing the contribution to $N_{\rm add}$ becomes independent of the index $\nu$ whereas for sufficiently small lattice spacing only $\nu=0$ contributes significantly.

\begin{center}
\begin{figure}[t!]
\includegraphics[scale=0.23]{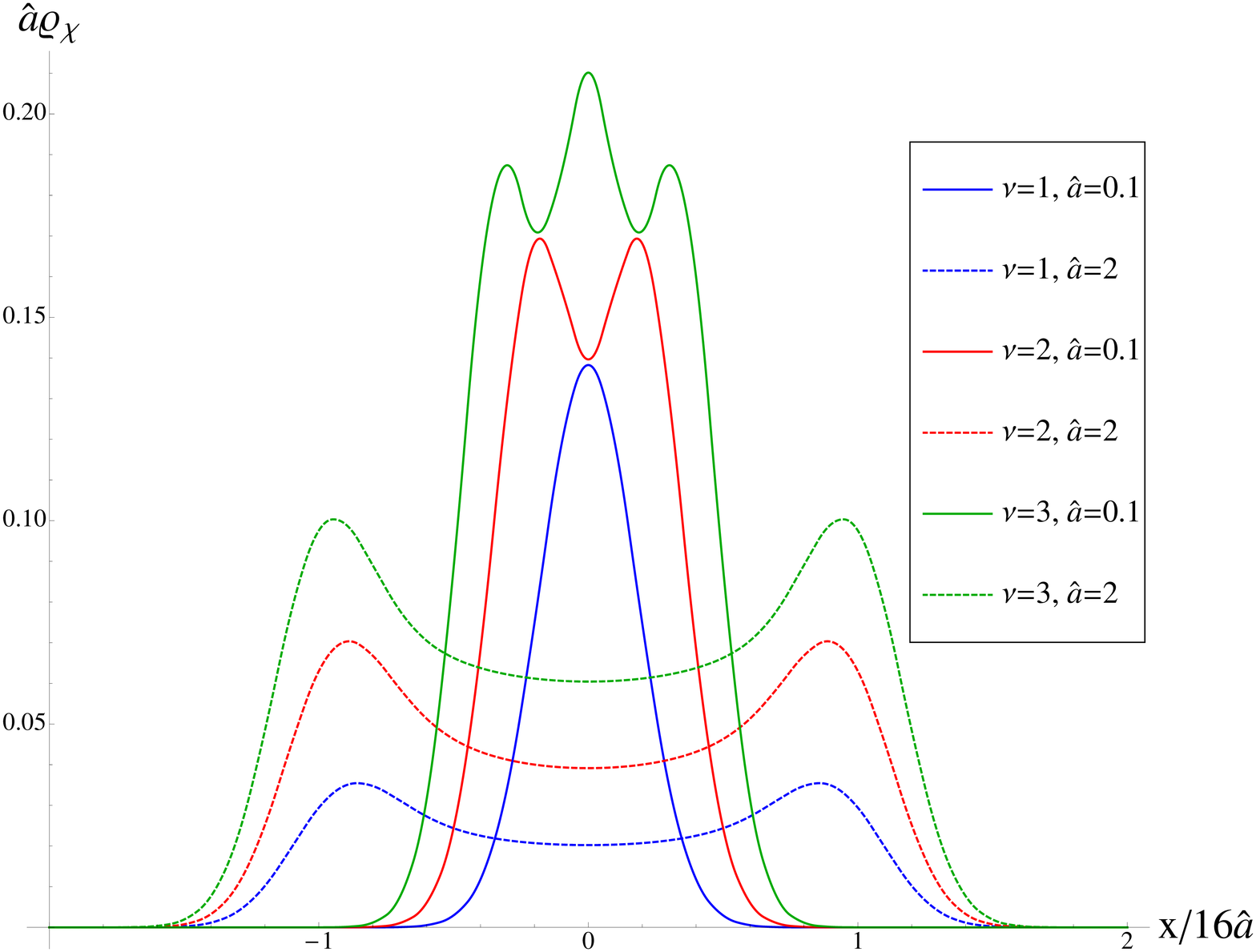}
\caption{For small lattice spacing (solid curves) the distribution 
$\rho_\chi$ is given by the GUE (See the legend for the values of the index
 and the lattice spacing). For $ \widehat{a} \gg 1$ (dashed curves) the shape becomes 
$\nu$-independent with two peaks at $\pm 8 \widehat{a}^2$   that 
behave as $1/\sqrt{(8\widehat{a}^2)^2-x^2} $ for $ \widehat{a} \gg1$.
}
\vspace*{-0.5cm}\label{fig4}
\includegraphics[scale=0.22]{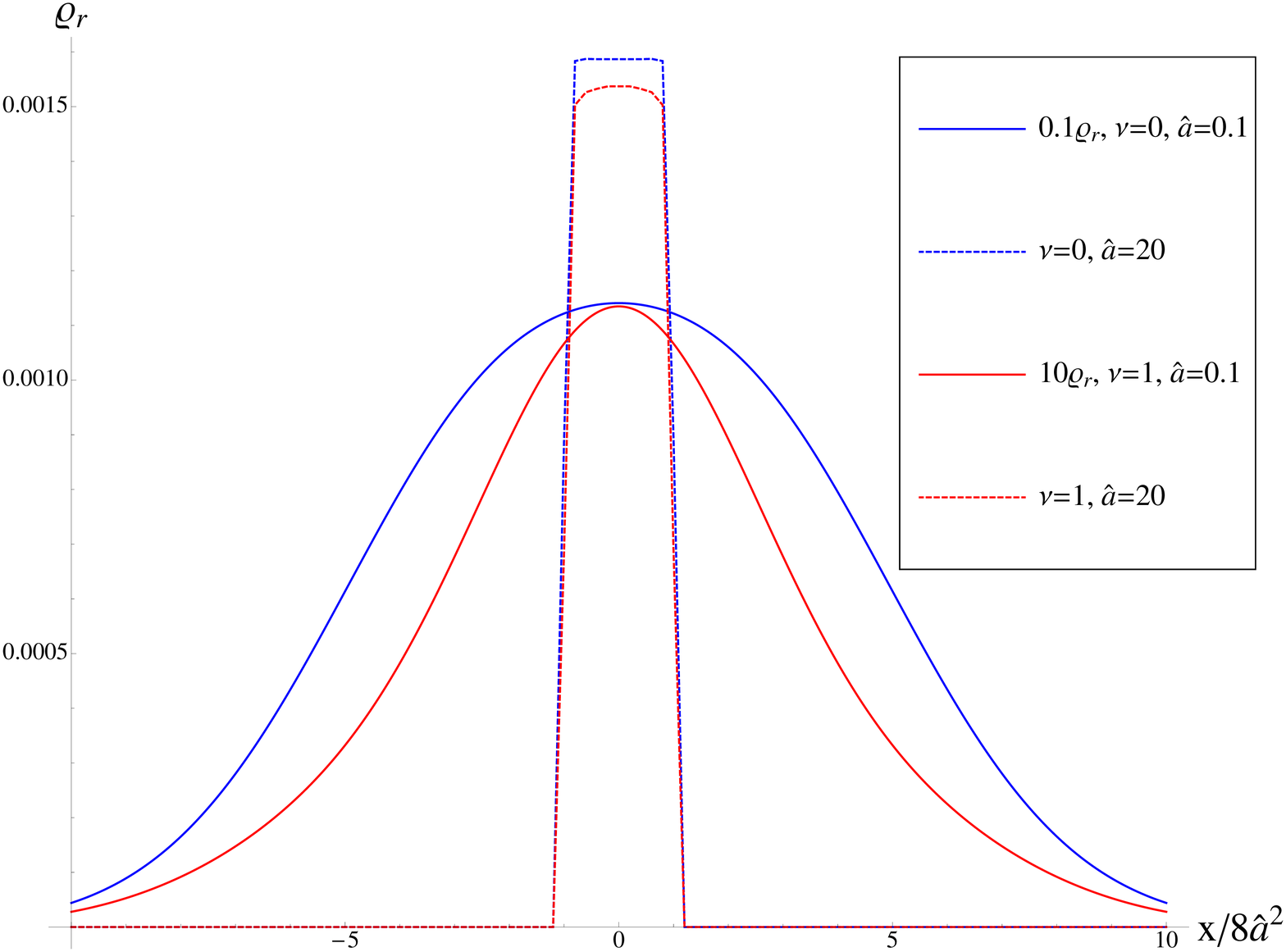}
\caption{The eigenvalue distribution $\rho_{\rm r}$ has a Gaussian shape for small $\widehat{a}$ (solid lines) but becomes box like with increasing lattice spacing (dashed curves). Notice that $\rho_{\rm r}$ for ($(\nu,\widehat{a})=(0,0.1)$) and ($(\nu,\widehat{a})=(1,0.1)$) is one order larger and one order smaller than shown in the diagram.}
\label{fig3}
\end{figure}
\end{center}

 For small lattice spacing, the distribution $\rho_{\rm r}$ has a Gaussian 
shape with a width of $2\widehat{a}$, and for  $\widehat{a}\gg 1$, it develops a plateau 
with sharp edges at $\pm 8\widehat{a}^2$, cf. Fig.~\ref{fig3}.
The height of $\rho_{\rm r}$ at the origin  scales 
like $\widehat{a}^{2\nu+1}$ for small lattice spacing 
and like $\widehat{a}^{-1}$ for large $\widehat{a}$.

The distribution of chirality over the real eigenvalues $\rho_\chi$ is shown in Fig.~\ref{fig4}.
 For small $\widehat{a}$ we observe the spectral density of 
 the $\nu$-dimensional Gaussian unitary ensemble. For large lattice spacing it
deforms into a curve with two peaks  at $\pm 8 \widehat{a}^2$ that up
to an overall normalization is independent of $\nu$ and evolves into inverse square root singularities for  $ \widehat a \to \infty$.

\vspace*{-0.95cm}\paragraph{Conclusions.}
Discretization effects become strong for $\widehat a \approx 0.5$. 
The oscillations of the spectral density in the continuum limit are no
longer visible while the density of the complex eigenvalues 
develops a plateau with a width of $16\widehat a^2$. In terms of
physical parameters, $\widehat{a} =  \widetilde{a} \sqrt{W_8 V}$, 
with $W_8$ a  low energy constant \cite{ADSV10b} and $V$ the volume of space time, 
we have the condition that $ \widetilde{a} \ll 1/\sqrt{W_8 V}$ to be close to the continuum limit.

In the regime of small lattice spacing, $\widehat{a}\approx 0.1$, 
the width of the distribution
of the complex eigenvalues is given by $\sigma = 2 \widetilde{a} \sqrt {W_8/V} /\Sigma$
whereas the spacing of the projection of these eigenvalues onto the 
imaginary axis is equal to $ \Delta \lambda = \pi /\Sigma V$. We thus have
that $\sigma/\Delta \lambda =2\widehat a /\pi $, which allows us to extract
a numerical  value for $W_8$ from lattice simulations.

An important result is that the number of additional real modes 
is strongly suppressed for large $\nu$. This implies that for large 
volumes when most configurations have an index $|\nu|  >  0$, additional
real modes are not much of a problem for lattice QCD simulations with
Wilson fermions provided that $W_8 a^2 V \ll 1$.

\vspace*{-0.4cm}
\section*{Acknowledgements}

We thank Gernot Akemann and Kim Splittorff for helpful comments. MK is financially supported by the Alexander-von-Humboldt Foundation. JV and SZ are supported by
U.S. DOE Grant No. DE-FG-88ER40388.


\end{document}